%% file: patharticle.tex
\documentclass[a4paper,10pt]{lncs}
\usepackage{amsmath}
\usepackage{pstricks}
\usepackage{graphicx}
\usepackage[latin1]{inputenc}

\usepackage{tikz}
\usetikzlibrary{backgrounds}
\usetikzlibrary{shapes}
\usetikzlibrary{calc}

\usepackage{fancyvrb}

\title{Back-translation for discovering distant protein homologies}
\author{Marta G\^irdea \and Laurent No\'e \and Gregory Kucherov\thanks{On leave in J.-V.Poncelet Lab, Moscow, Russia}}
\institute{INRIA Lille - Nord Europe, LIFL/CNRS, Universit\'e Lille 1, 59655 Villeneuve d'Ascq, France}

\begin{document}

\maketitle

\begin{abstract}

Frameshift mutations in protein-coding DNA sequences produce a drastic
change in the resulting protein sequence, which prevents classic
protein alignment methods from revealing the proteins' common
origin. Moreover, when a large number of substitutions are
additionally involved
in the divergence, the homology detection becomes difficult even at the DNA
level. To cope with this situation, we propose a novel method to infer
distant homology relations of two proteins, that accounts for frameshift and point
mutations that may have affected the coding sequences. We design a dynamic programming alignment algorithm over
memory-efficient graph representations of the complete set of putative
DNA sequences of each protein, with the goal of determining the two
putative DNA sequences which have the best scoring alignment under a powerful scoring system designed to reflect the most
probable evolutionary process. This allows us to uncover
evolutionary information that is not captured by traditional alignment
methods, which is confirmed by biologically significant examples. 
\end{abstract}

% Keywords: back-translation, protein, acyclic graph, dynamic programming, hidden homology, translation-dependent substitution score

% =============================================================================================
% INTRO
%
\section{Introduction}\label{sec:intro}

In protein-coding DNA sequences, frameshift mutations (insertions or deletions of one or more bases) can alter the translation reading frame, affecting all the amino acids encoded from that point forward. Thus, frameshifts produce a drastic change in the resulting protein sequence, preventing any similarity to be visible at the amino acid level.

When the coding DNA sequence is relatively well conserved, the similarity remains detectable at the DNA level, by DNA sequence alignment, as reported in several papers, including \cite{raes2005fdp,okamura2006fan,harrison2007fdh,hahn2005inh}.

However, the divergence often involves additional base substitutions. It has been shown~\cite{grantham1980ccu,shepherd1981mdr,guigo1999dcc} that, in coding DNA, there is a base compositional bias among codon positions, that does not apply when the translation reading frame is changed. Hence, after a reading frame change, a coding sequence is likely to undergo base substitutions leading to a composition that complies with this bias. Amongst these substitutions, synonymous mutations (usually occurring on the third position of the codon) are more likely to be accepted by natural selection, since they are silent with respect to the gene's product. If, in a long evolutionary time, a large number of codons in one or both sequences are affected by these changes, the sequence may be altered to such an extent that the common origin becomes difficult to observe by direct DNA comparison.

In this paper, we address the problem of finding distant protein homologies, in particular when the primary cause of the divergence is a frameshift. We achieve this by computing the best alignment of DNA sequences that encode the target proteins. This approach relies on the idea that synonymous mutations cause mismatches in the DNA alignments that can be avoided when all the sequences with the same translation are explored, instead of just the known coding DNA sequences. This allows the algorithm to search for an alignment by dealing only with non-synonymous mutations and gaps.

We designed and implemented an efficient method for aligning putative coding DNA sequences, which builds expressive alignments between hypothetical nucleotide sequences that can provide some information about the common ancestral sequence, if such a sequence exists. We perform the analysis on memory-efficient graph representations of the complete set of putative DNA sequences for each protein, described in Section~\ref{subsec:data-preprocessing}. The proposed method, presented in Section ~\ref{subsec:alignment-algo}, consists of a dynamic programming alignment algorithm that computes the two putative DNA sequences that have the best scoring alignment under an appropriate scoring system (Section \ref{subsec:scores}) designed to reflect the actual evolution process from a codon-oriented perspective.

While the idea of finding protein relations by frameshifted DNA alignments is not entirely new, as we will show in Section~\ref{sec:related-work} in a brief related work overview, Section \ref{sec:analysis} -- presenting tests performed on artificial data -- demonstrates the efficiency of our scoring system for distant sequences. Furthermore, we validate our method on several pairs of sequences known to be encoded by overlapping genes, and on some published examples of frameshifts resulting in functional proteins. We briefly present these experiments in Section \ref{sec:experiments}, along with a study of a protein family whose members present high dissimilarity on a certain interval. The paper is concluded in Section \ref{sec:conclusion}.

% =============================================================================================
% RELATED WORK
%
\section{Related Work}\label{sec:related-work}

The idea of using knowledge about coding DNA when aligning amino acid sequences has been explored in several papers.

A {\em non-statistical approach} for analyzing the homology and the ``genetic semihomology'' in protein sequences was presented in~\cite{leluk1998naa,leluk2000nsa}.
Instead of using a statistically computed scoring matrix, amino acid similarities are scored according to the complexity of the substitution process at the DNA level, depending on the number and type (transition/transversion) of nucleotide changes that are necessary for replacing one amino acid by the other.
This ensures a differentiated treatment of amino acid substitutions at different positions of the protein sequence, thus avoiding possible rough approximations resulting from scoring them equally, based on a classic scoring matrix.
The main drawback of this approach is that it was not designed to cope with frameshift mutations..

Regarding {\em frameshift mutation discovery}, many studies \cite{raes2005fdp,okamura2006fan,harrison2007fdh,hahn2005inh} preferred the plain BLAST~\cite{altschul1990bla,altschul25gba} alignment approach: BLASTN on DNA and mRNA, or BLASTX
on mRNA and proteins, applicable only when the DNA sequences are sufficiently similar. BLASTX programs, although capable of insightful results thanks to the six frame translations, have the limitation of not being able to transparently manage frameshifts that occur inside the sequence, for example by reconstructing an alignment from pieces obtained on different reading frames.

An interesting approach for {\em handling frameshifts at the protein level} was developed in~\cite{pellegrini1999sfe}. Several substitution matrices were designed for aligning amino acids encoded on different reading frames, based on nucleotide pair matches between respective codons.
This idea has the advantage of being easy to use with any classic protein alignment tool. However, it lacks flexibility in gap positioning.

On the subject of {\em aligning coding DNA in presence of frameshift errors}, some related ideas were presented in~\cite{arvestad1264acd,arvestad2000absa}. The author proposed to search for protein homologies by aligning their {\em sequence graphs} (data structures similar to the ones we describe in Section~\ref{subsec:data-preprocessing}). The algorithm tries to align pairs of codons, possibly incomplete since gaps of size 1 or 2 can be inserted at arbitrary positions. The score for aligning two such codons is computed as the maximum substitution score of two amino acids that can be obtained by translating them. This results in a complex, time costly dynamic programming method that basically explores all the possible translations. In Section~\ref{subsec:alignment-algo}, we present an algorithm addressing the same problem, more efficient since it aligns symbols, not codons, and more flexible with respect to scoring functions. Additionally, we propose to use a scoring system relying on codon evolution rather than amino acid translations, since we believe that, in frameshift mutation scenarios, the information provided by DNA sequence dynamics is more relevant than amino acid similarities.

% =============================================================================================
% OUR METHOD
%
\section{Our approach to distant protein relation discovery}\label{sec:approach}

The problem of inferring homologies between distantly related proteins, whose divergence is the result of frameshifts and point mutations, is approached in this paper by determining the best pairwise alignment between two DNA sequences that encode the proteins.

Given two proteins $P_A$ and $P_B$, the objective is to find a pair of DNA sequences, $D_A$ and $D_B$, such that $translation(D_A) = P_A$ and $translation(D_B) = P_B$, which produce the best pairwise alignment under a given scoring system.

The alignment algorithm (described in Section~\ref{subsec:alignment-algo}) incorporates a gap penalty that limits the number of frameshifts allowed in an alignment, to comply with the observed frequency of frameshifts in a coding sequence's evolution. The scoring system (Section~\ref{subsec:scores}) is based on possible mutational patterns of the sequences. 
This leads to reducing the false positive rate and focusing on alignments that are more likely to be biologically significant. 

% ---------------------------------------------------------------------------------------------
% Data structures
%
\subsection{Data structures}\label{subsec:data-preprocessing}

An explicit enumeration and pairwise alignment of all the putative DNA sequences is not an option, since their number increases exponentially with the protein's length\footnote{With the exception of $M$ and $W$, which have a single corresponding codon, all amino acids are encoded by 2, 3, 4 or 6 codons.}.
Therefore, we represent the protein's ``back-translation''  (set of possible source DNAs) as a directed acyclic graph, whose size depends linearly on the length of the protein, and where a path represents one putative sequence.

As illustrated in Figure~\ref{fig:compressed-graph1}(a), the graph is organized as a sequence of length $3n$ where $n$ is the length of the protein sequence. At each position $i$ in the graph, there is a group of nodes, each representing a possible nucleotide that can appear at position $i$ in at least one of the putative coding sequences. Two nodes at consecutive positions are linked by arcs if and only if they are either consecutive nucleotides of the same codon, or they are respectively the third and the first base of two consecutive codons. No other arcs exist in the graph. 

Note that 
in the implementation, the number of nodes is reduced by using the IUPAC nucleotide codes.
If the amino acids composing a protein sequence are non-ambiguous, only 4 extra nucleotide symbols -- $R$, $Y$, $H$ and $N$ -- are necessary for their back-translation. In this condensed representation, 
the number of ramifications in the graph is substantially reduced, as illustrated by Figure \ref{fig:compressed-graph1}. More precisely, the only amino acids with ramifications in their back-translation are amino acids $R$, $L$ and $S$, each encoded by 6 codons with different prefixes.

% ---------------------------------------------------------
% Algorithm
\subsection{Alignment algorithm}\label{subsec:alignment-algo}
We use a dynamic programming method, similar to the Smith-Waterman algorithm, extended to data structures described in Section \ref{subsec:data-preprocessing} and equipped with gap related restrictions.

\begin{figure}[ht]
\begin{minipage}[c]{0.53\textwidth}
\setlength{\unitlength}{5mm}
 \input{graphics/matrix-example}
\end{minipage}
\begin{minipage}[c]{0.46\textwidth}
 {\em $M[i, j]$ is a ``cell'' of $M$ corresponding to position $i$ of the first graph and position $j$ of the second graph.}

{\em $M[i, j]$ contains entries $(\alpha_A, \alpha_B)$ corresponding to pairs of nodes occurring in the first graph at position $i$, and in the second graph at position $j$, respectively.}
\vskip2em
\end{minipage}
\caption{Example of dynamic programming matrix $M$.}
\label{fig:alignment-matrix}
\end{figure}
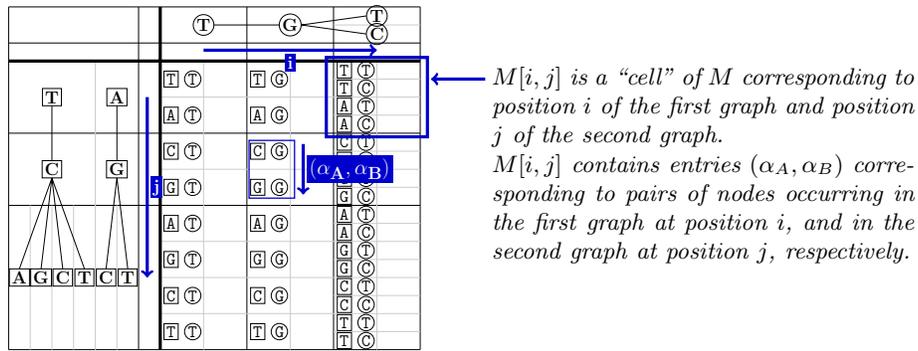

Given the input graphs $G_A$ and $G_B$ obtained by back-translating proteins $P_A$ and $P_B$, the algorithm finds the best scoring local alignment between two DNA sequences comprised in the back-translation graphs (illustrated in Figure~\ref{fig:alignment-ex}). The alignment is built by filling each entry $M[i, j, (\alpha_A,\alpha_B)]$ of a dynamic programming matrix $M$, where $i$ and $j$ are positions of the first and second graph respectively, and  $(\alpha_A, \alpha_B)$ is a pair of nodes that can be found in $G_A$ at position $i$, and in $G_B$ at position $j$, respectively.
An example is given in Figure \ref{fig:alignment-matrix}.

\begin{figure}
\begin{minipage}[c]{0.37\textwidth} 
\begin{center}
 \input{graphics/graphs-example}
\end{center}
\caption{Example of fully represented (a) and condensed (b) back-translation graph for the amino acid sequence YSH.}
\label{fig:compressed-graph1}
\end{minipage}
\hspace{0.5cm}
\begin{minipage}[c]{0.57\textwidth}
\begin{center}
  \input{graphics/alignment-example}
\end{center}
\caption{Alignment example. A path (corresponding to a putative DNA sequence) was chosen from each graph so that the match/mismatch ratio is maximized.}
\label{fig:alignment-ex}
\end{minipage}
\end{figure}
%-------------------------------------------------------------------------------------------------------------------------------------------

The dynamic programming algorithm begins with a classic local alignment initialization (0 at the top and left borders), followed by the recursion step described in equation~\eqref{eq-algo-rec2g}. The partial alignment score from each cell $M[i, j, (\alpha_A,\alpha_B)]$ is computed as the maximum of 6 types of values:
\begin{itemize}
 \item[(a)] $0$ (similarly to the classic Smith-Waterman algorithm, only non-negative scores are considered for local alignments).
 \item[(b)] the substitution score of symbols $(\alpha_A,\alpha_B)$, denoted $score(\alpha_A, \alpha_B)$, added to the score of the best partial alignment ending in $M[i-1, j-1]$, provided that the partially aligned paths contain $\alpha_A$ on position $i$ and $\alpha_B$ on position $j$ respectively; this condition is ensured by restricting the entries of $M[i-1, j-1]$ to those labeled with symbols that precede $\alpha_A$ and $\alpha_B$ in the graphs.
 \item[(c)] the cost $singleGapPenalty$ of a frameshift (gap of size 1 or extension of a gap of size 1) in the first sequence, added to the score of the best partial alignment that ends in a cell $M[i, j-1, (\alpha_A, \beta_B)]$, provided that $\beta_B$ precedes $\alpha_B$ in the second graph; this case is considered only if the number of allowed frameshifts on the current path is not exceeded, or a gap of size~1 is extended.
 \item[(d)] the cost of a frameshift in the second sequence, added to a partial alignment score defined as above.
 \item[(e)] the cost $tripleGapPenalty$ of removing an entire codon from the first sequence, added to the score of the best partial alignment ending in a cell $M[i, j-3, (\alpha_A, \beta_B)]$.%, with no restrictions imposed to $\beta_B$.
 \item[(f)] the cost of removing an entire codon from the second sequence, added to the score of the best partial alignment ending in a cell $M[i-3, j, (\beta_A, \alpha_B)]$
\end{itemize}

We adopted a non-monotonic gap penalty function, which favors insertions and deletions of full codons, and does not allow a large number of frameshifts -- very rare events, usually eliminated by natural selection. As can be seen in equation~(\ref{eq-algo-rec2g}), two particular kinds of gaps are considered: {\bf i) frameshifts} -- gaps of size 1 or 2, with high penalty, whose number in a local alignment can be limited, and {\bf ii) codon skips} -- gaps of size 3 which correspond to the insertion or deletion of a whole codon.
\begin{equation}\label{eq-algo-rec2g}
\begin{array}{l}
M[i, j, (\alpha_A,\alpha_B)] = \\
max \left \{
\begin{array}{llc}
0 & & \rm{(a)} \\
M[i-1, j-1, (\beta_A, \beta_B)] + score(\alpha_A, \alpha_B), & \beta_{k} \in pred(\alpha_{k}); & \rm{(b)}\\
\left(M[i, j-1, (\alpha_A, \beta_B)] + singleGapPenalty\right), & \beta_B \in pred(\alpha_B); & \rm{(c)}\\
\left(M[i-1, j, (\beta_A, \alpha_B)] + singleGapPenalty\right), & \beta_A \in pred(\alpha_A); & \rm{(d)}\\
\left(M[i, j-3, (\alpha_A, \beta_B)] + tripleGapPenalty\right), & j \ge 3 & \rm{(e)}\\
\left(M[i-3, j, (\beta_A, \alpha_B)] + tripleGapPenalty\right), & i \ge 3 & \rm{(f)}\\
\end{array}
\right .\\
\end{array}
\end{equation}

%---------------------------------------------------------------------------------------------
% Scores
\subsection{Translation-dependent scoring function}\label{subsec:scores}

In this section, we present a new translation-dependent scoring system suitable for our alignment algorithm. The scoring scheme we designed incorporates information about possible mutational patterns for coding sequences, based on a codon substitution model, with the aim of filtering out alignments between sequences that are unlikely to have common origins.

Mutation rates have been shown to vary within genomes, under the influence of several factors, including neighbor bases~\cite{blake1992inn}. Consequently, a model where all base mismatches are equally penalized is oversimplified, and ignores possibly precious information about the context of the substitution.

With the aim of retracing the sequence's evolution and revealing which base substitutions are more likely to occur within a given codon,
our scoring system
targets pairs of triplets $(\alpha, p, a)$, were $\alpha$ is a nucleotide, $p$ is its position in the codon, and $a$ is the amino acid encoded by that codon, thus differentiating various contexts of a substitution. There are 99 valid triplets out of the total of 240 hypothetical combinations.

Pairwise alignment scores are computed for all possible pairs of valid triplets  $(t_1, t_2) = ((\alpha_1, p_1, a_1), (\alpha_2, p_2, a_2))$ as a classic log-odds ratio:
\begin{equation}
score(t_1, t_2) = \lambda \log{\frac{f_{t_1 t_2}}{b_{t_1 t_2}}}\label{eq:score}
\end{equation}
where $f_{t_1 t_2}$ is the frequency of the $t_1 \leftrightarrow t_2$ substitution in related sequences, and $b_{t_1 t_2}=p(t_1)p(t_2)$ is the background probability.

% Method overview
In order to obtain the foreground probabilities $f_{t_i t_j}$, we will consider the following scenario: two proteins are encoded on the same DNA sequence, on different reading frames; at some point, the sequence was duplicated and the two copies diverged independently; we assume that the two coding sequences undergo, in their independent evolution, synonymous and non-synonymous point mutations, or full codon insertions and removals.

The insignificant amount of available real data that fits our hypothesis does not allow classical, statistical computation of the foreground and background probabilities. Therefore, instead of doing statistics on real data directly, we will rely on codon frequency tables and codon substitution models.

We assume that codon substitutions in our scenarios can be modeled by a Markov model presented in~\cite{kosiol2007ecm}\footnote{Another, more advanced codon substitution model, targeting sequences with overlapping reading frames, is proposed and discussed in~\cite{pedersen2001drm}. It does not fit our scenario, because it is designed for overlapping reading frames, where a mutation affects both translated sequences, while in our case the sequences become at one point independent and undergo mutations independently.
} which specifies the relative instantaneous substitution rate from codon $i$ to codon $j$ as:
\begin{equation}
Q_{ij} = \left\{\begin{array}{ll}
          0 & \mbox{if $i$ or $j$ is a stop codon, or} \\
            & \mbox{if $i \rightarrow j$ requires more than 1 nucleotide substitution,} \\
          \pi_j & \mbox{if $i \rightarrow j$ is a synonymous transversion,}  \\
          \pi_j \kappa &  \mbox{if $i \rightarrow j$ is a synonymous transition,} \\
          \pi_j \omega &  \mbox{if $i \rightarrow j$ is a nonsynonymous transversion,} \\
          \pi_j \kappa \omega &  \mbox{if $i \rightarrow j$ is a nonsynonymous transition.} \\
         \end{array} \right.
\end{equation}
for all $i \neq j$. Here, the parameter $\omega$ represents the nonsynonymous-synonymous rate ratio, $\kappa$ the transition-transversion rate ratio, and $\pi_j$ the equilibrium frequency of codon $j$. As in all Markov models of sequence evolution, absolute rates are found by normalizing the relative rates to a mean rate of 1 at equilibrium, that is, by enforcing $\sum_i\sum_{j\neq i} \pi_i Q_{ij} = 1$ and completing the instantaneous rate matrix $Q$ by defining $Q_{ii} = -\sum_{j\neq i} Q_{ij}$ to give a form in which the transition probability matrix is calculated as $P({\theta}) = e^{\theta Q}$~\cite{lio1998mme}. Evolutionary times $\theta$ are measured in expected number of nucleotide substitutions per codon.

% Foreground
With this codon substitution model, $f_{t_i t_j}$ can be deduced in several steps. Basically, we first need to identify all  pairs of codons with a common subsequence, that have a perfect semi-global alignment (for instance, codons $CAT$ and $ATG$ satisfy this condition, having the common subsequence $AT$; this example is further explained below). We then assume that the codons from each pair undergo independent evolution, according to the codon substitution model. For the resulting codons, we compute, based on all possible original codon pairs, $p((\alpha_i, p_i, c_i), (\alpha_j, p_j, c_j))$ -- the probability that nucleotide $\alpha_i$, situated on position $p_i$ of codon $c_i$, and nucleotide $\alpha_j$, situated on position $p_j$ of codon $c_j$ have a common origin (equation \eqref{eq:fg-fs-all}). From these, we can immediately compute, as shown by equation \eqref{eq:fg-aa}, $p((\alpha_i, p_i, a_i), (\alpha_j, p_j, a_j))$, corresponding in fact to the foreground probabilities $f_{t_i t_j}$, where $t_i = (\alpha_i, p_i, a_i)$ and $t_j = (\alpha_j, p_j, a_j)$.

In the following, $\bf{p(c_1 \stackrel{\theta}{\rightarrow} c_2)}$ stands for the probability of the event {\em codon $c_1$ mutates into codon $c_2$ in the evolutionary time $\theta$}, and is given by $P_{c_1, c_2}({\theta})$.

$\bf{c_1[interval_1]\equiv c_2[interval_2]}$ states that codon $c_1$ restricted to the positions given by  $interval_1$ is a sequence identical to $c_2$ restricted to $interval_2$.
This is equivalent to having a word $w$ obtained by ``merging'' the two codons. For instance, if $c_1 = CAT$ and $c_2 = ATG$, with their common substring being placed in $interval_1 = [2..3]$ and $interval_2=[1..2]$ respectively, $w$ is $CATG$.

Finally, $\bf{p(c_1[interval_1]\equiv c_2[interval_2])}$ is the probability to have $c_1$ and $c_2$, in the relation described above, which we compute as the probability of the word $w$ obtained by ``merging'' the two codons. This function should be symmetric,
it should depend on the codon distribution, and the probabilities of all the words $w$ of a given length should sum to~1.
However, since we consider the case where the same DNA sequence is translated on two different reading frames, one of the two translated sequences would have an atypical composition. Consequently, the probability of a word $w$ is computed as if the sequence had the known codon composition when translated on the reading frame imposed by the first codon, or on the one imposed by the second. This hypothesis can be formalized as:
\begin{equation}
p(w)=p({\scriptstyle \mbox{\small $w$ on $rf_{1}$ OR $w$ on $rf_{2}$}}) = p^{rf_{1}}(w) + p^{rf_{2}}(w) - p^{rf_{1}}(w) \cdot  p^{rf_{2}}(w)
\end{equation}
where $p^{rf_{1}}(w)$ and $p^{rf_{2}}(w)$ are the probabilities of the word $w$ in the reading frame imposed by the position of the first and second codon, respectively. This is computed as the products of the probabilities of the codons and codon pieces that compose the word $w$ in the established reading frame.
In the previous example, the probabilities of $w =CATG$ in the first and second reading frame are:
\vskip-1em
$$ p^{rf_1}(CATG) = p(CAT) \cdot p(G**) = p(CAT) \cdot \sum_{c: c\mbox{ \scriptsize starts with } G}{p(c)} $$
\vskip-1em
$$ p^{rf_2}(CATG) = p(**C) \cdot p(ATG) = \sum_{c: c\mbox{ \scriptsize ends with } C}{p(c)} \cdot p(ATG) $$

The values of $p((\alpha_i, p_i, c_i), (\alpha_j, p_j, c_j))$ are computed as:
\vskip-1em
\begin{equation}
\sum_{\stackrel{c'_i, c'_j:c'_i[interval_i]\equiv c'_j[interval_j]}{p_i \in interval_i, p_j \in interval_j}}{\!\!\!\!\!\!\!\!\!\!\!\!\!\!\!p(c'_i[interval_i]\equiv c'_j[interval_j]) \cdot p(c'_i \stackrel{\theta}{\rightarrow} c_i) \cdot p(c'_j \stackrel{\theta}{\rightarrow} c_j)}
 \label{eq:fg-fs-all}
\end{equation}
from which obtaining the {\bf foreground probabilities} is straightforward:
\vskip-1.5em
\begin{equation}
f_{t_i t_j} = p((\alpha_i, p_i, a_i), (\alpha_j, p_j, a_j)) = \sum_{\stackrel{c_i \mbox{\scriptsize encodes } a_i,}{\stackrel{c_j \mbox{\scriptsize encodes } a_j}{}}}{p((\alpha_i, p_i, c_i), (\alpha_j, p_j, c_j))}
 \label{eq:fg-aa}
\end{equation}

% Background
The {\bf background probabilities} of $(t_i, t_j)$, $b_{t_i t_j}$, can be simply expressed as the probability of the two symbols appearing independently in the sequences:
\vskip-0.75em
\begin{equation}
b_{t_i t_j} = b_{(\alpha_i, p_i, a_i), (\alpha_j, p_j, a_j)} = \sum_{\stackrel{c_i \mbox{\scriptsize encodes } a_i,}{\stackrel{c_j \mbox{\scriptsize encodes } a_j}{}}}\pi_{c_i}\pi_{c_j}
 \label{eq:bg-fs-all-simple}
\end{equation}

% Scores for compressed graphs
\paragraph{Substitution matrix for ambiguous symbols}

From matrices built as explained above, the versions that use IUPAC ambiguity codes for nucleotides (as proposed in the final paragraph of \ref{subsec:data-preprocessing}) can be computed: the score of pairing two ambiguous symbols is the maximum over all substitution scores for  all pairs of nucleotides from the respective sets.

% Statistics
\paragraph{Score evaluation}
The score significance is estimated according to the Gumbel distribution, where the parameters $\lambda$ and $K$ are computed with the method described in~\cite{altschul2001esp,olsen1999rae}. Since the forward alignment and the reverse complementary alignment are two independent cases with different score distributions, two parameter pairs, $\lambda_{fw}, K_{fw}$ and $\lambda_{rc}, K_{rc}$ are computed and used in practice.

% =====================================================================================
% Analysis

\section{Validation}\label{sec:analysis}

To validate the translation-dependent scoring system we designed in the previous section, we tested it on an artificial data set consisting in 96 pairs of protein sequences of average length 300. Each pair was obtained by translating a randomly generated DNA sequence on two different reading frames. Both sequences in each pair were then mutated independently, according to codon mutation probability matrices corresponding to each of the evolutionary times 0.01, 0.1, 0.3, 0.5, 0.7, 1.0, 1.5, 2.00 (measured in average number of mutations per codon).

To this data set we applied four variants of alignment algorithms:
i) classic alignment of DNA sequences using classic base substitution scores and affine gap penalties;
ii) classic alignment of DNA sequences using a translation-dependent scoring scheme designed in Section~\ref{subsec:scores};
iii) alignment of back-translation graphs (Section~\ref{subsec:alignment-algo}) using classic base substitution scores and affine gap penalties;
iv) alignment of back-translation graphs using a translation-dependent scoring scheme. 
For the tests involving translation-dependent scores, we used scoring functions corresponding to  evolutionary times from 0.30 to 1.00.

Table \ref{table:validation} 
briefly shows the e-values of the scores obtained with each setup when aligning sequence pairs with various evolutionary distances.
While all variants perform well for highly similar sequences, we can clearly deduce the ability of the translation-dependent scores to help the algorithm build significant alignments between sequences that underwent important changes.
\input{data/validation-results}
\vskip-2em
The resulting alignments reveal that, even after many mutations, the transla-tion-dependent scores manage to recover large parts of the original shared sequence, by correctly aligning most positions.
On the other hand, with classic match/mismatch scores, the algorithm usually fails to find these common zones. Moreover, due to the large number of mismatches, the alignment has a low score, comparable to scores that can be obtained for randomly chosen sequences. This makes it difficult to establish whether the alignment is biologically meaningful or it was obtained by chance. This issue is solved by the translation-dependent scores by uneven substitution penalties, according to the codon mutation models.

We conclude that the usage of translation-dependent scores makes the
algorithm more robust, able to detect the common origins even after the sequences underwent many modifications, and also able to filter out alignments where the nucleotide pairs match by pure chance and not due to evolutionary relations.

% =====================================================================================
% Results
%
\vspace{-1mm}
\section{Experimental results}\label{sec:experiments}

% --------------------------------------------------------------------------------------
% Validation
%
\vspace{-1mm}
\subsection{Tests on known overlapping and frameshifted genes}
\vspace{-1mm}
We tested the method on pairs of proteins known to be encoded by overlapping genes in viral genomes (phage X174 and Influenza A) and in E.coli plasmids, as well as on the newly identified overlapping genes {\em yaaW} and {\em htgA} from E.coli K12~\cite{delaye2008ong}. 
In all cases, we obtained perfect identification of gene overlaps with simple substitution scores and with translation-dependent scoring matrices corresponding to low evolutionary distances (at most 1 mutation per codon ).
Translation-dependent scoring matrices of higher evolutionary distances favor, in some (rare) cases, substitutions instead of matches within the alignment. This is a natural consequence of increasing the codon's chance to mutate, and it illustrates the importance of choosing a score matrix corresponding to the real evolutionary distance.
Our method was also able to detect, directly on the protein sequences, the frameshifts resulting in functional proteins reported in~\cite{raes2005fdp,okamura2006fan,harrison2007fdh,hahn2005inh}.

% --------------------------------------------------------------------------------------
% Experiment
\vspace{-1.5mm}
\subsection{New divergence scenarios for orthologous proteins}
\label{sec:human-orphan-genes}
\vspace{-1mm}
In this section we discuss the application of our method to FMR1NB (Fragile X mental retardation 1 neighbor protein) family. The Ensembl database~\cite{ensembl} provides 23 members of this family, from mammalian species, including human,  mouse, dog and cow. Their multiple alignment, provided by Ensembl, shows high dissimilarity on the first part (100 amino acids approximately), and good conservation 
on the rest of the sequence. We apply our alignment algorithm on proteins from several organisms, where the complete sequence is available.

We performed our experiments with translation-dependent scoring matrices corresponding to 0.3, 0.5 and 0.7 mutations per codon. Given that, in our scenario (presented in section~\ref{subsec:scores}), the divergence is applied on two reading frames, this implies an overall mutation rate of 0.6, 1.0 and 1.4 mutations per codon respectively. Thus, the mutation rate per base reflected by our scores is less than 0.5, which is approximately the nucleotide substitution rate for mouse relative to human~\cite{clamp2007cdp}.
The number of allowed frameshifts was limited to 3. The gap penalties were set in all cases to -20 for codon indels, -20 for size 1 gaps and -5 for the extension of size 1 gaps (size 1 and size 2 gaps correspond to frameshifts). These choices were made so that the penalty for codon indels is higher than the average penalty for 3 substitutions.

\input{data/alignment-fmr1nb}%
Figure \ref{fig:fmr1-al} presents a fragment of the alignment obtained on the FMR1NB proteins of human (gene ID ENSG00000176988) and mouse (gene ID ENSMUSG00000062170).
The algorithm finds a frameshift near the 100th amino acid, managing
to align the initial part of the proteins at the DNA level. Similar
frameshifted alignments are obtained for human vs. cow and human
vs. dog, while alignments between proteins of primates do not contain
frameshifts. The consistency of the frameshift position in these
alignments supports the evidence of a frameshift event that might have
occurred in the primate lineage. 

If confirmed, this frameshift would have modified the first topological
domain and the first transmembrane domain of the product
protein. Interestingly, the FMR1NB gene occurs nearby the Fragile X mental
retardation 1 gene (FMR1), involved in the corresponding genetic
disease~\cite{oostra2001fxg}.

% =====================================================================================
% Conclusions
%
\vspace{-1.5mm}
\section{Conclusions}\label{sec:conclusion}
\vspace{-1mm}
In this paper, we addressed the problem of finding distant protein homologies, in particular affected by frameshift events, from a codon evolution perspective. We search for protein common origins by implicitly aligning all their putative coding DNA sequences, stored in efficient data structures called back-translation graphs. Our approach relies on a dynamic programming alignment algorithm for these graphs, which involves a non-monotonic gap penalty that handles differently frameshifts and full codon indels.
We designed a powerful translation-dependent scoring function for nucleotide pairs, based on codon substitution models, whose purpose is to reflect the expected dynamics of coding DNA sequences.

The method was shown to perform better than classic alignment on artificial data, obtained by mutating independently, according to a codon substitution model coding sequences translated with a frameshift. Moreover, it successfully detected published frameshift mutation cases resulting in functional proteins.

We then described an experiment involving homologous mammalian proteins that showed little conservation at the amino acid level on a large region, and provided possible frameshifted alignments obtained with our method, that may explain the divergence.
As illustrated by this example, the proposed method should allow to better explain a high divergence of homologous proteins and to help to establish new homology relations between genes with unknown origins.

An implementation of our method is available at \url{http://bioinfo.lifl.fr/path/}.
\bibliographystyle{splncs}%
\bibliography{sources}%
\end{document}

%% file: graphics/graphs-example.tex
\begin{tikzpicture}[tight background,
    aa/.style={font=\huge,anchor=mid},
    symbol/.style={anchor=mid,inner sep=0pt},
    arc/.style={},
    arrow/.style={double distance=7pt,->,decorate,very thick,draw=black!50},
    scale=0.6, transform shape
    ]
\path[draw,white] (0pt,0pt) rectangle (160pt, 100pt);
\begin{scope}[transform canvas={scale=1}]
 \coordinate (SeqStart) at (12pt, 12pt);
 \coordinate (arrowLen) at (80pt, 0pt);
 \coordinate (arrowHalf) at (0pt, 32pt);
 \coordinate (arrowWidth) at (12pt, 0pt);
 \coordinate (sep) at (6pt, 0pt);
 \coordinate (dy) at (0pt, 6pt);

 % AA sequence
 \node (YSH) [aa, anchor=east, text centered] at ($ (SeqStart) $){YSH};

 % ``Back-translation" arrow
 \node (arrow1) [anchor=west, draw,shape=single arrow,single arrow tip angle=130,shape border rotate=270,rotate=90,minimum width=64pt,minimum height=10pt] at ($ (YSH.east) + (sep) - (arrowHalf) $) {Back-translation};

 % Back-translation graph
  \coordinate (GStart) at ($ (arrow1.east) - (arrowHalf) + (arrowWidth) + (sep) + 5*(dy)$);
 % Y
 \node (T1Y) [symbol, anchor=west] at (GStart) {T};
 \node (A2Y) [symbol, anchor=west] at ($ (T1Y.east) + (sep) $) {A};
 \node (C3Y) [symbol, anchor=west] at ($ (A2Y.east) + (sep) + (dy) $) {C};
 \node (T3Y) [symbol, anchor=west] at ($ (A2Y.east) + (sep) - (dy) $) {T};
 \draw (T1Y) -- (A2Y) (A2Y) -- (C3Y) (A2Y) -- (T3Y);

 % S
 \node (T1S) [symbol, anchor=west] at ($ (C3Y.east) + (sep) $) {T};
 \node (A1S) [symbol, anchor=west] at ($ (T3Y.east) + (sep) $) {A};
 \node (C2S) [symbol, anchor=west] at ($ (T1S.east) + (sep) $) {C};
 \node (G2S) [symbol, anchor=west] at ($ (A1S.east) + (sep) $) {G};

 \node (A3S) [symbol, anchor=west] at ($ (C2S.east) + (sep) + 6*(dy) $) {A};
 \node (C3S) [symbol, anchor=west] at ($ (C2S.east) + (sep) + 4*(dy) $) {C};
 \node (G3S) [symbol, anchor=west] at ($ (C2S.east) + (sep) + 2*(dy) $) {G};
 \node (T3S) [symbol, anchor=west] at ($ (C2S.east) + (sep) $) {T};

 \node (C3S') [symbol, anchor=west] at ($ (G2S.east) + (sep) $) {C};
 \node (T3S') [symbol, anchor=west] at ($ (G2S.east) + (sep) - 2*(dy) $) {T};

 \draw (T1S) -- (C2S) (C2S) -- (A3S) (C2S) -- (C3S) (C2S) -- (G3S) (C2S) -- (T3S) (A1S) -- (G2S) (G2S) -- (C3S') (G2S) -- (T3S');

 % Y-S arcs
 \draw (C3Y) -- (T1S) (C3Y) -- (A1S) (T3Y) -- (T1S) (T3Y) -- (A1S);

 % h
 \node (C1H) [symbol, anchor=west] at ($ (T3S.east) + (sep) - (dy) $) {C};
 \node (A2H) [symbol, anchor=west] at ($ (C1H.east) + (sep) $) {A};
 \node (C3H) [symbol, anchor=west] at ($ (A2H.east) + (sep) + (dy) $) {C};
 \node (T3H) [symbol, anchor=west] at ($ (A2H.east) + (sep) - (dy) $) {T};
 \draw (C1H) -- (A2H) (A2H) -- (C3H) (A2H) -- (T3H);

 \node (a) [aa, anchor=west] at ($ (T3H) + (sep) + (dy) $) {(a)};

 % S-H arcs
 \draw (C1H) -- (A3S) (C1H) -- (C3S) (C1H) -- (G3S) (C1H) -- (T3S) (C1H) -- (C3S') (C1H) -- (T3S');

 \node (arrow2) [anchor=west, draw=black!50,shape=double arrow,double arrow tip angle=120,rotate=90,minimum width=14pt,minimum height=24pt] at ($ (C2S.south) - 8*(dy) $) {};

 % small graph
  \coordinate (CStart) at ($ (arrow1.east) - (arrowHalf) + (arrowWidth) + (sep) - 5*(dy)$);
 \node (T1Yc) [symbol, anchor=west] at ($ (CStart) $) {T};
 \node (A2Yc) [symbol, anchor=west] at ($ (T1Yc.east) + (sep) $) {A};
 \node (Y3Yc) [symbol, anchor=west] at ($ (A2Yc.east) + (sep) $) {Y};
 \node (T1Sc) [symbol, anchor=west] at ($ (Y3Yc.east) + (sep) + (dy) $) {T};
 \node (C2Sc) [symbol, anchor=west] at ($ (T1Sc.east) + (sep) $) {C};
 \node (N3Sc) [symbol, anchor=west] at ($ (C2Sc.east) + (sep) $) {N};
 \node (A1Sc) [symbol, anchor=west] at ($ (Y3Yc.east) + (sep) - (dy)  $) {A};
 \node (G2Sc) [symbol, anchor=west] at ($ (A1Sc.east) + (sep) $) {G};
 \node (Y3Sc) [symbol, anchor=west] at ($ (G2Sc.east) + (sep) $) {Y};
 \node (C1Hc) [symbol, anchor=west] at ($ (N3Sc.east) + (sep)  - (dy) $) {C};
 \node (A2Hc) [symbol, anchor=west] at ($ (C1Hc.east) + (sep) $) {A};
 \node (Y3Hc) [symbol, anchor=west] at ($ (A2Hc.east) + (sep) $) {Y};
 \draw (T1Yc) -- (A2Yc) -- (Y3Yc) -- (T1Sc) -- (C2Sc) -- (N3Sc) -- (C1Hc) -- (A2Hc) -- (Y3Hc) (Y3Yc) -- (A1Sc) -- (G2Sc) -- (Y3Sc) -- (C1Hc);
 \node (b) [aa, anchor=west] at ($ (Y3Hc) + (sep) $) {(b)};
\end{scope}
\end{tikzpicture}

%% file: graphics/alignment-example.tex
\begin{tikzpicture}[tight background,
    aa/.style={font=\huge,anchor=mid},
    symbol/.style={anchor=mid,inner sep=0pt},
    othersymbol/.style={anchor=mid,inner sep=0pt,black!50},
    arc/.style={},
    otherarc/.style={draw=black!50},
    scale=0.6, transform shape
    ]
\path[draw,white] (0pt,-23pt) rectangle (250pt, 75pt);
\begin{scope}[transform canvas={scale=1}]
 \coordinate (SeqStart) at (12pt, 12pt);
 \coordinate (rowSep) at (0pt, 40pt);
 \coordinate (arrowLen) at (80pt, 0pt);
 \coordinate (sep) at (6pt, 0pt);
 \coordinate (dy) at (0pt, 6pt);
 \coordinate (ATarrowWidth) at (0, 0);
 \coordinate (graphsWidth) at (4.3cm, 0);

 % -----------------------------------------------
 % AA sequences
 \node (AGN) [aa, anchor=east, text centered] at ($ (SeqStart) $){AGN:};
 \node (QET) [aa, anchor=east, text centered] at ($ (SeqStart) - (rowSep) $){QET:};

 % -----------------------------------------------
 % Back-translation grapH 1
 \coordinate (G1Start) at ($ (AGN.east) + (ATarrowWidth) $);
 % A
 \node (C1A) [symbol, anchor=west] at (G1Start) {C};
 \node (C2A) [symbol, anchor=west] at ($ (C1A.east) + (sep) $) {C};
 \node (A3A) [symbol, anchor=west] at ($ (C2A.east) + (sep) + 3*(dy) $) {A};
 \node (C3A) [symbol, anchor=west] at ($ (C2A.east) + (sep) + (dy) $) {C};
 \node (G3A) [symbol, anchor=west] at ($ (C2A.east) + (sep) - (dy) $) {G};
 \node (T3A) [symbol, anchor=west] at ($ (C2A.east) + (sep) - 3*(dy) $) {T};
 \draw (C1A) -- (C2A) (C2A) -- (A3A) (C2A) -- (C3A) (C2A) -- (G3A) (C2A) -- (T3A);

 % G
 \node (G1G) [symbol, anchor=west] at ($ (T3A.east) + (sep) + 3*(dy) $) {G};
 \node (G2G) [symbol, anchor=west] at ($ (G1G.east) + (sep) $) {G};
 \node (A3G) [symbol, anchor=west] at ($ (G2G.east) + (sep) + 3*(dy) $) {A};
 \node (C3G) [symbol, anchor=west] at ($ (G2G.east) + (sep) + (dy) $) {C};
 \node (G3G) [symbol, anchor=west] at ($ (G2G.east) + (sep) - (dy) $) {G};
 \node (T3G) [symbol, anchor=west] at ($ (G2G.east) + (sep) - 3*(dy) $) {T};
 \draw (G1G) -- (G2G) (G2G) -- (A3G) (G2G) -- (C3G) (G2G) -- (G3G) (G2G) -- (T3G);

 % A-G arcs
 \draw  (G1G) -- (A3A) (G1G) -- (C3A) (G1G) -- (G3A) (G1G) -- (T3A);

 % N
 \node (A1N) [symbol, anchor=west] at ($ (T3G.east) + (sep) + 3*(dy)$) {A};
 \node (A2N) [symbol, anchor=west] at ($ (A1N.east) + (sep) $) {A};
 \node (C3N) [symbol, anchor=west] at ($ (A2N.east) + (sep) + (dy) $) {C};
 \node (T3N) [symbol, anchor=west] at ($ (A2N.east) + (sep) - (dy) $) {T};
 \draw (A1N) -- (A2N) (A2N) -- (C3N) (A2N) -- (T3N);

 % G-N arcs
 \draw (A1N) -- (A3G) (A1N) -- (C3G) (A1N) -- (G3G) (A1N) -- (T3G);

 % -----------------------------------------------
 % Back-translation grapH 2
 \coordinate (G2Start) at ($ (QET.east) + (ATarrowWidth) $);

  % Q
 \node (C1Q) [symbol, anchor=west] at (G2Start) {C};
 \node (A2Q) [symbol, anchor=west] at ($ (C1Q.east) + (sep) $) {A};
 \node (A3Q) [symbol, anchor=west] at ($ (A2Q.east) + (sep) + (dy) $) {A};
 \node (G3Q) [symbol, anchor=west] at ($ (A2Q.east) + (sep) - (dy) $) {G};
 \draw (C1Q) -- (A2Q) (A2Q) -- (A3Q) (A2Q) -- (G3Q);

  % E
 \node (G1E) [symbol, anchor=west] at ($ (G3Q.east) + (sep) + (dy)$) {G};
 \node (A2E) [symbol, anchor=west] at ($ (G1E.east) + (sep) $) {A};
 \node (A3E) [symbol, anchor=west] at ($ (A2E.east) + (sep) + (dy) $) {A};
 \node (G3E) [symbol, anchor=west] at ($ (A2E.east) + (sep) - (dy) $) {G};
 \draw (G1E) -- (A2E) (A2E) -- (A3E) (A2E) -- (G3E);

 % Q-E arcs
 \draw (G1E) -- (A3Q) (G1E) -- (G3Q);

 % T
 \node (A1T) [symbol, anchor=west] at ($ (G3E.east) + (sep) + (dy)$) {A};
 \node (C2T) [symbol, anchor=west] at ($ (A1T.east) + (sep) $) {C};
 \node (A3T) [symbol, anchor=west] at ($ (C2T.east) + (sep) + 3*(dy) $) {A};
 \node (C3T) [symbol, anchor=west] at ($ (C2T.east) + (sep) + (dy) $) {C};
 \node (G3T) [symbol, anchor=west] at ($ (C2T.east) + (sep) - (dy) $) {G};
 \node (T3T) [symbol, anchor=west] at ($ (C2T.east) + (sep) - 3*(dy) $) {T};
 \draw (A1T) -- (C2T) (C2T) -- (A3T) (C2T) -- (C3T) (C2T) -- (G3T) (C2T) -- (T3T);

 % E-T arcs
 \draw (A1T) -- (A3E) (A1T) -- (G3E);

 % -----------------------------------------------
 % ``ALIGNMENT" arrow
\node (arrow2) [anchor=north,shape=single arrow,single arrow tip angle=130,shape border rotate=270,rotate=90,minimum width=24pt,minimum height=24pt,fill=blue!20] at ($ (AGN.east) - 0.5*(rowSep) + (ATarrowWidth) + (graphsWidth) $) {Alignment};

 % -----------------------------------------------
 % [ALIGNMENT] Back-translation grapH 1
\coordinate (G1AStart) at ($ (arrow2.south) + 0.5*(dy) $);
 % A
 \node (C1A) [othersymbol, anchor=west] at (G1AStart) {C};
 \node (C2A) [symbol, anchor=west] at ($ (C1A.east) + (sep) $) {C};
 \node (A3A) [symbol, anchor=west] at ($ (C2A.east) + (sep) $) {A};
 \node (C3A) [othersymbol, anchor=west] at ($ (C2A.east) + (sep) + 2*(dy) $) {C};
 \node (G3A) [othersymbol, anchor=west] at ($ (C2A.east) + (sep) + 4*(dy) $) {G};
 \node (T3A) [othersymbol, anchor=west] at ($ (C2A.east) + (sep) + 6*(dy) $) {T};

 % G
 \node (G1G) [symbol, anchor=west] at ($ (A3A.east) + (sep) $) {G};
 \node (G2G) [symbol, anchor=west] at ($ (G1G.east) + (sep) $) {G};
 \node (A3G) [symbol, anchor=west] at ($ (G2G.east) + (sep) $) {A};
 \node (C3G) [othersymbol, anchor=west] at ($ (G2G.east) + (sep) + 2*(dy) $) {C};
 \node (G3G) [othersymbol, anchor=west] at ($ (G2G.east) + (sep) + 4*(dy) $) {G};
 \node (T3G) [othersymbol, anchor=west] at ($ (G2G.east) + (sep) + 6*(dy) $) {T};

 % N
 \node (A1N) [symbol, anchor=west] at ($ (A3G.east) + (sep) $) {A};
 \node (A2N) [symbol, anchor=west] at ($ (A1N.east) + (sep) $) {A};
 \node (C3N) [symbol, anchor=west] at ($ (A2N.east) + (sep) $) {C};
 \node (T3N) [othersymbol, anchor=west] at ($ (A2N.east) + (sep) + 2*(dy) $) {T};

 % Chosen path arcs
 \draw (C2A) -- (A3A) -- (G1G) -- (G2G) -- (A3G) -- (A1N) -- (A2N) -- (C3N);
 % other arcs
 \draw [otherarc] (C1A) -- (C2A) (C2A) -- (C3A) (C2A) -- (G3A) (C2A) -- (T3A) (G2G) -- (C3G) (G2G) -- (G3G) (G2G) -- (T3G) (A2N) -- (T3N);
 \draw [otherarc]                (G1G) -- (C3A) (G1G) -- (G3A) (G1G) -- (T3A) (A1N) -- (C3G) (A1N) -- (G3G) (A1N) -- (T3G)               ;

% -----------------------------------------------
 % [ALIGNMENT]Back-translation grapH 2
 \coordinate (G2AStart) at ($ (C2A.south) - (dy) $);

  % Q
 \node (C1Q) [symbol, anchor=north] at (G2AStart) {C};
 \node (A2Q) [symbol, anchor=west] at ($ (C1Q.east) + (sep) $) {A};
 \node (A3Q) [othersymbol, anchor=west] at ($ (A2Q.east) + (sep) -2*(dy) $) {A};
 \node (G3Q) [symbol, anchor=west] at ($ (A2Q.east) + (sep) $) {G};

  % E
 \node (G1E) [symbol, anchor=west] at ($ (G3Q.east) + (sep) $) {G};
 \node (A2E) [symbol, anchor=west] at ($ (G1E.east) + (sep) $) {A};
 \node (A3E) [symbol, anchor=west] at ($ (A2E.east) + (sep) $) {A};
 \node (G3E) [othersymbol, anchor=west] at ($ (A2E.east) + (sep) - 2*(dy) $) {G};

 % T
 \node (A1T) [symbol, anchor=west] at ($ (A3E.east) + (sep) $) {A};
 \node (C2T) [symbol, anchor=west] at ($ (A1T.east) + (sep) $) {C};
 \node (A3T) [othersymbol, anchor=west] at ($ (C2T.east) + (sep) $) {A};
 \node (C3T) [othersymbol, anchor=west] at ($ (C2T.east) + (sep) - 2*(dy) $) {C};
 \node (G3T) [othersymbol, anchor=west] at ($ (C2T.east) + (sep) - 4*(dy) $) {G};
 \node (T3T) [othersymbol, anchor=west] at ($ (C2T.east) + (sep) - 6*(dy) $) {T};

 % Chosen path arcs
 \draw (C1Q) -- (A2Q) -- (G3Q) -- (G1E) -- (A2E) -- (A3E) -- (A1T) -- (C2T);
 % other arcs
 \draw [otherarc] (A2Q) -- (A3Q)  (A2E) -- (G3E) (C2T) -- (A3T) (C2T) -- (C3T) (C2T) -- (G3T) (C2T) -- (T3T);
 \draw [otherarc] (G1E) -- (A3Q)  (A1T) -- (G3E);

\end{scope}
\end{tikzpicture}

%% file: graphics/matrix-example.tex
\newrgbcolor{highlight}{0. 0. .8}
\newrgbcolor{highlight2}{0. 0. .8}
\newrgbcolor{seqAcolor}{.0 .0 .0}
\newrgbcolor{seqBcolor}{.0 .0 .0}

\begin{tikzpicture}[tight background,x=1.5cm,y=-1.5cm,font=\ttfamily\huge,
    header/.style={font=\huge\bfseries}, 
    seq1/.style={draw=black,shape=rectangle},
    seq2/.style={draw=black,shape=circle, inner sep=1pt},
    scale=0.8, transform shape
]

\path[draw,white] (-14pt,-160pt) rectangle (100pt, 11pt);
\begin{scope}[transform canvas={scale=0.4}]
    \pgftransformshift{\pgfpoint{1cm}{1}}
    \draw [thick] (-1.2, 0) -- (10.2, 0) (-1.2, 1) -- (10.2, 1) (-1.2, 1.5) -- (10.2, 1.5) (-1.2, 3.5) -- (10.2,3.5)
          (-1.2, 5.5) -- (10.2, 5.5) (-1.2, 9.5) -- (10.2, 9.5);
    \draw [thick] (-1.2, 0) -- (-1.2, 9.5) (2.4, 0) -- (2.4, 9.5) (3, 0) -- (3, 9.5) (5.4, 0) -- (5.4, 9.5)
          (7.8, 0) -- (7.8, 9.5) (10.2, 0) -- (10.2, 9.5);

    \foreach \x in {1.2, 4.2, 6.6, 9}
      \draw[help lines, very thin, black!20] (\x, 1.5) -- (\x, 9.5);
    \foreach \x in {-0.6, 0, 0.6, 1.2, 1.8}
      \draw[help lines, very thin, black!20] (\x, 5.5) -- (\x, 9.5);
    \foreach \y in {2.5, 4.5, 6.5, 7.5, 8.5}
      \draw[help lines, very thin, black!20] (3, \y) -- (10.2, \y);
    \foreach \y in {0.5, 2, 3, 4, 5, 6, 7, 8, 9}
      \draw[help lines, very thin, black!20] (7.8, \y) -- (10.2, \y);

    \draw [line width=3pt] (-1.2, 1.5) -- (10.2, 1.5) (3, 0) -- (3, 9.5) ;

% First sequence

    \node (T1) [header] at (0.0, 2.5)[seqAcolor,seq1] {T};
    \node (A1) [header] at (1.8, 2.5)[seqAcolor,seq1] {A};
    \node (C2) [header] at (0.0, 4.5)[seqAcolor,seq1] {C};
    \node (G2) [header] at (1.8, 4.5)[seqAcolor,seq1] {G};
    \node (A3) [header] at (-0.9, 7.5)[seqAcolor,seq1] {A};
    \node (G3) [header] at (-0.3, 7.5)[seqAcolor,seq1] {G};
    \node (C3) [header] at (0.3, 7.5)[seqAcolor,seq1] {C};
    \node (T3) [header] at (0.9, 7.5)[seqAcolor,seq1] {T};
    \node (C3') [header] at (1.5, 7.5)[seqAcolor,seq1] {C};
    \node (T3') [header] at (2.1, 7.5)[seqAcolor,seq1] {T};
    \draw[thick, seqAcolor] (T1) -- (C2) (A1) -- (G2)
          (C2.south) -- (A3.north) (C2.south) -- (G3.north) (C2.south) -- (C3.north) (C2.south) -- (T3.north)
          (G2.south) -- (C3'.north) (G2.south) -- (T3'.north);

% Second sequence

    \node (T'1) [header] at (4.2, 0.5)[seqAcolor,seq2] {T};
    \node (G'2) [header] at (6.6, 0.5)[seqAcolor,seq2] {G};
    \node (T'3) [header] at (9, 0.25)[seqAcolor,seq2] {T};
    \node (C'3) [header] at (9, 0.75)[seqAcolor,seq2] {C};
    \draw[thick, seqBcolor] (T'1) -- (G'2)
          (G'2.east) -- (T'3.west) (G'2.east) -- (C'3.west);

% Matix
    % 1 1
    \node (T1T'1-l) at (3.3, 2)[seqAcolor,seq1] {T};
    \node (T1T'1-r) at (3.9, 2)[seqAcolor,seq2] {T};
    \node (A1T'1-l) at (3.3, 3)[seqAcolor,seq1] {A};
    \node (A1T'1-r) at (3.9, 3)[seqAcolor,seq2] {T};
    % 1 2
    \node (T1G'2-l) at (5.7, 2)[seqAcolor,seq1] {T};
    \node (T1G'2-r) at (6.3, 2)[seqAcolor,seq2] {G};
    \node (A1G'2-l) at (5.7, 3)[seqAcolor,seq1] {A};
    \node (A1G'2-r) at (6.3, 3)[seqAcolor,seq2] {G};
    % 1 3
    \node (T1T'3-l) at (8.1, 1.75)[seqAcolor,seq1] {T};
    \node (T1T'3-r) at (8.7, 1.75)[seqAcolor,seq2] {T};
    \node (T1C'3-l) at (8.1, 2.25)[seqAcolor,seq1] {T};
    \node (T1C'3-r) at (8.7, 2.25)[seqAcolor,seq2] {C};
    \node (A1T'3-l) at (8.1, 2.75)[seqAcolor,seq1] {A};
    \node (A1T'3-r) at (8.7, 2.75)[seqAcolor,seq2] {T};
    \node (A1C'3-l) at (8.1, 3.25)[seqAcolor,seq1] {A};
    \node (A1C'3-r) at (8.7, 3.25)[seqAcolor,seq2] {C};

    % 2 1
    \node (C2T'1-l) at (3.3, 4)[seqAcolor,seq1] {C};
    \node (C2T'1-r) at (3.9, 4)[seqAcolor,seq2] {T};
    \node (G2T'1-l) at (3.3, 5)[seqAcolor,seq1] {G};
    \node (G2T'1-r) at (3.9, 5)[seqAcolor,seq2] {T};
    % 2 2
    \node (C2G'2-l) at (5.7, 4)[seqAcolor,seq1] {C};
    \node (C2G'2-r) at (6.3, 4)[seqAcolor,seq2] {G};
    \node (G2G'2-l) at (5.7, 5)[seqAcolor,seq1] {G};
    \node (G2G'2-r) at (6.3, 5)[seqAcolor,seq2] {G};
    % 2 3
    \node (C2T'3-l) at (8.1, 3.75)[seqAcolor,seq1] {C};
    \node (C2T'3-r) at (8.7, 3.75)[seqAcolor,seq2] {T};
    \node (C2C'3-l) at (8.1, 4.25)[seqAcolor,seq1] {C};
    \node (C2C'3-r) at (8.7, 4.25)[seqAcolor,seq2] {C};
    \node (G2T'3-l) at (8.1, 4.75)[seqAcolor,seq1] {G};
    \node (G2T'3-r) at (8.7, 4.75)[seqAcolor,seq2] {T};
    \node (G2C'3-l) at (8.1, 5.25)[seqAcolor,seq1] {G};
    \node (G2C'3-r) at (8.7, 5.25)[seqAcolor,seq2] {C};

    % 3 1
    \node (A3T'1-l) at (3.3, 6)[seqAcolor,seq1] {A};
    \node (A3T'1-r) at (3.9, 6)[seqAcolor,seq2] {T};
    \node (G3T'1-l) at (3.3, 7)[seqAcolor,seq1] {G};
    \node (G3T'1-r) at (3.9, 7)[seqAcolor,seq2] {T};
    \node (C3T'1-l) at (3.3, 8)[seqAcolor,seq1] {C};
    \node (C3T'1-r) at (3.9, 8)[seqAcolor,seq2] {T};
    \node (T3T'1-l) at (3.3, 9)[seqAcolor,seq1] {T};
    \node (T3T'1-r) at (3.9, 9)[seqAcolor,seq2] {T};
    % 3 2
    \node (A3G'2-l) at (5.7, 6)[seqAcolor,seq1] {A};
    \node (A3G'2-r) at (6.3, 6)[seqAcolor,seq2] {G};
    \node (G3G'2-l) at (5.7, 7)[seqAcolor,seq1] {G};
    \node (G3G'2-r) at (6.3, 7)[seqAcolor,seq2] {G};
    \node (C3G'2-l) at (5.7, 8)[seqAcolor,seq1] {C};
    \node (C3G'2-r) at (6.3, 8)[seqAcolor,seq2] {G};
    \node (T3G'2-l) at (5.7, 9)[seqAcolor,seq1] {T};
    \node (T3G'2-r) at (6.3, 9)[seqAcolor,seq2] {G};
    % 3 3
    \node (A3T'3-l) at (8.1, 5.75)[seqAcolor,seq1] {A};
    \node (A3T'3-r) at (8.7, 5.75)[seqAcolor,seq2] {T};
    \node (A3C'3-l) at (8.1, 6.25)[seqAcolor,seq1] {A};
    \node (A3C'3-r) at (8.7, 6.25)[seqAcolor,seq2] {C};
    \node (G3T'3-l) at (8.1, 6.75)[seqAcolor,seq1] {G};
    \node (G3T'3-r) at (8.7, 6.75)[seqAcolor,seq2] {T};
    \node (G3C'3-l) at (8.1, 7.25)[seqAcolor,seq1] {G};
    \node (G3C'3-r) at (8.7, 7.25)[seqAcolor,seq2] {C};
    \node (C3T'3-l) at (8.1, 7.75)[seqAcolor,seq1] {C};
    \node (C3T'3-r) at (8.7, 7.75)[seqAcolor,seq2] {T};
    \node (C3C'3-l) at (8.1, 8.25)[seqAcolor,seq1] {C};
    \node (C3C'3-r) at (8.7, 8.25)[seqAcolor,seq2] {C};
    \node (T3T'3-l) at (8.1, 8.75)[seqAcolor,seq1] {T};
    \node (T3T'3-r) at (8.7, 8.75)[seqAcolor,seq2] {T};
    \node (T3C'3-l) at (8.1, 9.25)[seqAcolor,seq1] {T};
    \node (T3C'3-r) at (8.7, 9.25)[seqAcolor,seq2] {C};
    \draw [->, line width=3pt, highlight] (4.2, 1.2) -- node [below] {\colorbox{highlight}{\textcolor{white}{\bf i}}} (9, 1.2);
    \draw [->, line width=3pt, highlight] (2.64, 2.5) -- node [right] {\colorbox{highlight}{\textcolor{white}{\bf j}}} (2.64, 7.5);
    \draw [very thick, highlight] (5.46, 3.7) rectangle (6.72, 5.3);
    \draw [->, line width=3pt, highlight] (6.96, 3.8) -- node [right] {\colorbox{highlight}{\textcolor{white}{\Huge$\mathbf{(\alpha_A,\alpha_B)}$}}} (6.96, 5.2);

    \draw [line width=3pt, highlight] (10.4, 1.4) rectangle (7.6, 3.6);
    \draw [->, line width=3pt, highlight] (12, 2) -- (10.5, 2);

\end{scope}
\end{tikzpicture}

%% file: data/validation-results.tex
\newcommand{\pwd}[1]{$\cdot 10^{-#1}$}
\newcommand{\ord}[1]{$10^{-#1}$}
\begin{table}
\begin{center}
\begin{tabular}[htb]{|cc|c|c|c|c|c|c|c|c|c|}
\hline
\multicolumn{3}{|c|}{~}                                                            & \multicolumn{8}{|c|}{\bf Evolutionary distance between the aligned inputs} \\
\hline
      & {\bf Scores$^{(*)}$} & {\bf Input type} &~~{\bf 0.01}~~&~~{\bf 0.10}~~&~~{\bf 0.30}~~&~~{\bf 0.50}~~&~~{\bf 0.70}~~&~~{\bf 1.00}~~&~~{\bf 1.50}~~&~~{\bf 2.00}~~\\
\hline
\hline
              & TDS 0.30     & graphs           & \ord{179} & \ord{171} & \ord{149}  & \ord{121} & \ord{109} & \ord{83} & \ord{61} & \ord{37}  \\
              &              & known DNAs       & \ord{152} & \ord{136} & \ord{110}  & \ord{76}  & \ord{54}  & \ord{21} & \ord{6}  & 1.00      \\
\hline
              & TDS 0.50     & graphs           & \ord{166} & \ord{156} & \ord{140}  & \ord{118} & \ord{107} & \ord{85} & \ord{55} & \ord{34}  \\
              &              & known DNAs       & \ord{140} & \ord{128} & \ord{105}  & \ord{75}  & \ord{61}  & \ord{34} & \ord{6}  & \ord{1}   \\
\hline
              & TDS 0.70     & graphs           & \ord{153} & \ord{145} & \ord{130}  & \ord{113} & \ord{102} & \ord{83} & \ord{57} & \ord{51}  \\
              &              & known DNAs       & \ord{130} & \ord{120} & \ord{101}  & \ord{76}  & \ord{64}  & \ord{42} & \ord{13} & \ord{7}   \\
\hline
              & TDS 1.00     & graphs           & \ord{137} & \ord{131} & \ord{118}  & \ord{104} & \ord{97}  & \ord{80} & \ord{59} & \ord{54}  \\
              &              & known DNAs       & \ord{117} & \ord{110} & \ord{93}   & \ord{70}  & \ord{65}  & \ord{46} & \ord{21} & \ord{8}  \\
\hline
              & classic      & graphs           & \ord{127} & \ord{24}  & \ord{12}   & \ord{11}  & \ord{7}   & \ord{5}  & \ord{3}  & \ord{2}  \\
              & scores       & known DNAs       & \ord{86}  & \ord{20}  & \ord{9}    & \ord{7}   & \ord{4}   & \ord{1}  & 1.00     & 1.00    \\
\hline 
\end{tabular}
\end{center}

\caption{Order of the e-values of the scores obtained by aligning artificially diverged pairs of proteins resulted from the translation of the same ancestral sequence on two reading frames.
$^{(*)}${\em TDS $<$evolutionary distance$>$} = translation-dependent scores; classic substitution scores: match = 3, transversion = -4, transition = -2.}
\label{table:validation}
\end{table}

%% file: data/alignment-fmr1nb.tex
\begin{figure}[t]
\vspace{-1mm}
{\tiny
\begin{verbatim}
>Q8N0W7|FMR1N_HUMAN[14, 644] / Q80ZA7|FMR1N_MOUSE[0, 623]

...L][S][Y][Y][L][C][S][G][S][S][Y][F][V][L][A][N][G][H][I][L][P][N][S][E][N][A][H][G][Q][S][L][E][E][D][S][A][L][E][A][
...TCTCTTATTACCTTTGTTCTGGATCCTCGTACTTCGTTTTAGCCAACGGGCACATACTGCCAAACTCCGAAAACGCGCACGGGCAGTCGCTGGAAGAGGATTCGGCGTTAGAGGCCT
...|||||||||||:|||||||||||:|||||||.|||.||||||.|   .||:|.|||||||||||||||.||.||.||||.:|||||:||    ||..|||.:||.||||||:|:.|.
... ++++++++++0+++++++++++0+++++++-+++--+++++++   --+-0-+++++++++++0+++-0+--0--+++-+++++++++    +++--++++++--0++++++---+
...TCTCTTATTACTTTTGTTCTGGACCCTCGTAATTCTTTTTAGAC---CGGTAAATACTGCCAAACTCCCAATACTCGCAGAGGCAGCCG----AATTGGAACCGCCGTTAGGGAACA
...][L][L][L][L][L][F][W][T][L][V][I][L][F][R][   P][V][N][T][A][K][L][P][I][L][A][E][A][A][    E][L][E][P][P][L][G][N][

L][L][N][F][F][F][P][T][T][C][N][L][R][E][N][Q][V][A][K][P][C][N][E][L][Q][D][L][S][E][S][E][C][L][R][H][K][C][C][F][...
TGTTAAACTTTTTTTTTCCAACAACGTGTAACCTAAGAGAAAATCAAGTAGCGAAGCCGTGTAATGAGCTGCAGGACTTATCAGAATCAGAATGTTTAAGGCACAAATGTTGTTTTT...
|||||:|||||||||||||||||:||||||.|.|||||||.||||||||||:|:.|.||||||||:|.|.||.|.||||||||||||||||||||||||:|.:|||||||||||:||...
0++++++++++++++++++++++++++++000--+++++++++++++++++++-----0++++++-----0--+00+++++++++++++++++++++++++---+++++++++++++...
TGTTAGACTTTTTTTTTCCAACAGCGTGTATCATAAGAGATAATCAAGTAGTGGTGGCGTGTAATAACCAGCCGTACTTATCAGAATCAGAATGTTTAAAGAGCAAATGTTGTTCTT...
M][L][D][F][F][F][P][T][A][C][I][I][R][D][N][Q][V][V][V][A][C][N][N][Q][P][Y][L][S][E][S][E][C][L][K][S][K][C][C][S][...
\end{verbatim}
\vskip-4em
}
\begin{tikzpicture}[remember picture,overlay]
\node [xshift=13.32cm,yshift=24.04cm,opacity=0.25] at (current page.south west)
[text width=0.15cm,text height=1.0cm,fill=blue!80,above right]
{};
\end{tikzpicture}
\caption{Human and mouse FMR1NB proteins, aligned using a translation-dependent matrix of evolutionary distance 0.7 (the sign of each substitution score appears on the fourth row). The size 4 gap corresponds to a frameshift that corrects the reading frame.
}\label{fig:fmr1-al}
\vspace{-4mm}
\end{figure}